\documentclass[epsfig,12pt]{article}
\usepackage{epsfig}
\usepackage{graphicx}

\usepackage{latexsym}
\usepackage{amsmath}
\usepackage{amssymb}
\usepackage{relsize}
\usepackage{geometry}
\def\beq{\begin{equation}}
\def\eeq{\end{equation}}
\def\beqn{\begin{eqnarray}}
\def\eeqn{\end{eqnarray}}

\newcommand{\ntwo}{${\mathcal N}=2\,$}
\newcommand{\none}{${\mathcal N}=1\,$}

\newcommand{\cpn}{CP$(N-1)\,$}

\newcommand{\pt}{\partial}

\newcommand{\gsim}{\lower.7ex\hbox{$
\;\stackrel{\textstyle>}{\sim}\;$}}
\newcommand{\lsim}{\lower.7ex\hbox{$
\;\stackrel{\textstyle<}{\sim}\;$}}

\newcommand{\bra}{\left\langle}
\newcommand{\ket}{\right\rangle}

\def\slashed#1{\setbox0=\hbox{$#1$}             
   \dimen0=\wd0                                 
   \setbox1=\hbox{/} \dimen1=\wd1               
   \ifdim\dimen0>\dimen1                        
      \rlap{\hbox to \dimen0{\hfil/\hfil}}      
      #1                                        
   \else                                        
      \rlap{\hbox to \dimen1{\hfil$#1$\hfil}}   
      /                                         
   \fi}                                        %

\begin{document}


\begin{titlepage}

\begin{flushright}
UMN-TH-2930/10,   FTPI-10/36 \\
ITEP-TH-57/10 \\ January 5, 2011
\end{flushright}

\vspace{1cm}

\begin{center}

{  \Large \bf  Confined Magnetic Monopoles in Dense QCD}

\vspace{7mm}

 {\large
 \bf   A.~Gorsky$^{\,a,b}$,  M.~Shifman$^{\,b,c}$ and \bf A.~Yung$^{\,\,b,d}$}

\vspace{3mm}

$^a$
{\it Theory Department, ITEP, Moscow, Russia}\\[1mm]
$^b${\it  William I. Fine Theoretical Physics Institute,
University of Minnesota,
Minneapolis, MN 55455, USA}\\[1mm]
$^{c}${\it Center for Theoretical Physics, Department of Physics, Massachusetts Institute of Technology, Cambridge, MA 02139, USA}\\[1mm]
$^{d}${\it Petersburg Nuclear Physics Institute, Gatchina, St. Petersburg
188300, Russia
}
\end{center}

\vspace{1cm}

\begin{center}

{\large \bf Abstract}
\vspace {5mm}
\end{center}

Non-Abelian strings exist in the
color-flavor locked phase of  dense QCD.
We show that kinks appearing in the world-sheet theory on these strings,
in the form of the kink-antikink bound pairs, are the magnetic monopoles ---  descendants
of the 't Hooft--Polyakov monopoles surviving in such a special form in dense QCD.
Our consideration is heavily based on analogies and inspiration coming
from  certain supersymmetric non-Abelian theories.
This is the first ever analytic demonstration that
objects unambiguously identifiable  as the magnetic monopoles are ``native"
to non-Abelian Yang--Mills theories (albeit our analysis extends only to the phase of the monopole
confinement and has nothing to say about their condensation).
Technically our demonstration becomes possible due to the fact that
low-energy dynamics of the non-Abelian strings in dense QCD is
  that of the  orientational zero modes. It is
 described by  an effective two-dimensional
$CP(2)$ model  on the string world sheet. The kinks in this model representing
confined magnetic monopoles are   in a highly quantum regime.

\end{titlepage}

\newpage

\section{Introduction}
\label{intro}

Despite a huge number of works attempting to treat
the monopole condensation in Yang--Mills theories
as the quark confinement mechanism, the very notion of
the magnetic monopoles remains obscure in QCD:
a clear-cut gauge invariant framework for their description and analysis is
still absent. This is in contradistinction with a remarkable progress
in supersymmetric Yang--Mills theories where in 1994 Seiberg and Witten analytically  proved,
for the first time ever, that the dual Meissner effect does take place in a certain model,
chromoelectric flux tubes do form,
and quark confinement ensues \cite{SW1}.
Further explorations in this area led people to such finds as the non-Abelian flux tubes
\cite{HT1,ABEKY}
and confined magnetic monopoles \cite{SYmon,HT2} in a well-defined
and fully controllable setting.

In this paper we demonstrate that
{\em  confined} non-Abelian magnetic monopoles can be identified in a well-defined manner
in high-density QCD.
The very issue of flux tubes
in high-density QCD \cite{fthdqcd2,fthdqcd3,gm,fthdqcd1}
is the result of crossbreeding of two recent developments:
the discovery of non-Abelian flux tubes in supersymmetric gauge theories \cite{HT1,ABEKY,SYmon,HT2}
(for a review
see  \cite{Trev,Jrev,SYrev,Trev2}) and the color-flavor locked (CFL) phases
in QCD with a nonvanishing chemical potential $\mu$ \cite{1,2}.
In principle, non-Abelian flux tubes in dense QCD can show up either in quark-gluon plasma
or neutron stars. Leaving aside experimental identification issues we argue that
sufficiently long flux tubes (strings) of this type
support kink-antikink pairs in stable ``mesonic" states. The kinks, in their turn,
can be shown to represent the magnetic monopoles in the confined phase.
By the confined phase we mean that (i) the monopoles cannot be disassociated
from the strings attached to them; and (ii)   the monopole-antimonopole pairs are confined along the string,
while the parent QCD {\em per se} is in the CFL phase. It is not ruled out that
dualizing this picture \'{a} la   \cite{Shifman:2009mb} we
could arrive at a description of {\em condensed} dyons or monopoles.
This latter aspect is left for a separate study.

Large-$\mu$ QCD exhibits a rich  phase structure of quark-gluon matter
at  weak coupling.  With high enough chemical potential $\mu\gg\Lambda$,
the theory supports color  superconductivity due to
the Cooper pair condensation of diquarks with the vanishing orbital
momentum (for  reviews see ~\cite{1,2} ). Several
superconducting phases are known which
differ by the residual symmetry as well as the structure of the
condensates. For instance, in the so-called 2SC phase
in which only $u$ and $d$ quarks of two colors pair up
\cite{2SC}, the strange quark density
plays no role, while at larger values of $\mu$
 the color-flavor locked (CFL) phase is realized in which all three
quark flavors  condense,  implying the complete
breaking of the the non-Abelian global flavor group \cite{CFL}.
In addition, all eight gauge bosons are Higgsed. At the same time,
a diagonal global SU(3) survives.

It is the CFL phase of dense QCD which is the subject of our studies.
Magnetic monopoles in this environment have been already mentioned
(in the negative sense)
 in
\cite{fthdqcd3}. However, our
conclusion --- the presence of well-defined states of magnetic monopoles in the
CFL phase of dense QCD  --- is opposite to what was
advocated in \cite{fthdqcd3}. The crucial additional element
 of our analysis  which was absent in \cite{fthdqcd3}
 is the existence of kinks on the world sheet of the non-Abelian strings that form in dense QCD.

 The paper is organized as follows. We briefly review the CFL phase of dense QCD
 (Section  \ref{cflp}) and the emergence of non-Abelian strings in this phase (Section  \ref{nasicflp}),
 formulate an effective Ginzburg--Landau  description (Section  \ref{gled}),
 and then proceed to the demonstration that kinks supported by
 the Ginzburg--Landau model are in fact distorted magnetic monopoles ---  descendants
of the 't Hooft--Polyakov monopoles surviving in such a special form in
the environment of dense QCD (Sections  \ref{osws}, \ref{kkmesons} and~\ref{moncfl}).
 Section \ref{tamrs} is devoted to perfecting
 the simplified Ginzburg--Landau model of Section  \ref{gled} to making it  more realistic. In particular, in this section we introduce a nonvanishing mass for the strange quark. Section \ref{thetad}
 describes possible effects due to the $\theta$ term. Finally, Section \ref{conccc}
 briefly summarizes our results.

\section{Color-Flavor-Locked Phase}
\label{cflp}

To begin with we briefly outline the structure of the CFL phase in dense QCD.
The one-gluon exchange interaction results in nontrivial diquark condensates
of the Cooper type
\beqn
&&
\bra X_{kC}\ket \propto \delta_{kC}\neq 0\,,\qquad \bra Y^{Ck}\ket \propto \delta^{kC}\neq 0\,,\nonumber\\[3mm]
&&
 X_{kC}= q^{iA}_\alpha\,  q^{jB\,\alpha}\,\varepsilon_{ijk}\,\varepsilon_{ABC}\,,\qquad
 Y^{Ck}= \tilde{q}_{iA\, \alpha}\,  \tilde{q}_{jB}^{\alpha}\,\varepsilon^{ijk}\,\varepsilon^{ABC}\,
 \label{odin}
\eeqn
where the small and capital Latin letters mark color and flavor, respectively,
$q_\alpha$ and $\tilde{q}_{\alpha}$ are left-handed spinors; the first one is in the triplet representation
of the color and flavor SU(3) groups, while the second one is in the antitriplet representation.
The Dirac spinor of each
flavor is composed of a pair $(q_\alpha ,\, \bar{\tilde{q}}^{\dot\alpha})$.
Needless to say, the complex-conjugate condensates
$\bra X^\dagger\ket \,, \,\,\,  \bra Y^\dagger\ket$ do not vanish as well.
In the chiral limit, when $m_{u,d,s}=0$, the QCD Lagrangian is globally invariant under two
chiral SU(3) symmetries. In addition, it is invariant under U(1)$_B$ and U(1)$_A$ where $B$ marks the
baryon number and $A$ stands for axial. U(1)$_A$ is anomalous; however, at
 large $\mu$ the impact of anomaly is small and can be treated as a correction.

The $X$ condensate transforms under
SU(3)$_C\times$SU(3)$_L$ while the $Y^\dagger$ condensate under SU(3)$_C\times$SU(3)$_R$
where SU(3)$_{L,R}$ are the global chiral groups associated with flavor. The condensates (\ref{odin}),
break all three groups; however, the diagonal vectorial SU(3)$_{C+F}$ obviously survives.
In addition, the $X$ and $Y$ condensates spontaneously break  U(1)$_B$ and U(1)$_A$.
Altogether we have 8+8+2 Goldstone bosons in the   limit $m_{u,d,s}=0$. The boson
coupled to U(1)$_A$ is, in fact, a quasi-Goldstone. Since the vectorial flavor SU(3) is unbroken,
all Goldstones fall into representations of SU(3): two octets and two singlets.
Dynamics in  the CFL phase enjoy the properties of  systems with
superconductivity and superfluidity (due to the
broken U(1)$_B$ symmetry).

Out of eighteen Goldstone bosons eight
degrees of freedom corresponding to the vector-like fluctuations
are eaten by gluons via the Higgs mechanism. Hence we end up with
two Goldstone mesons corresponding to the broken Abelian symmetries (U(1)$_B$ and U(1)$_A$)  and
eight pseudoscalar Goldstones coupled to the axial SU(3) currents.\footnote{In supersymmetric theories
{\em all} would-be Goldstones are eaten up
by the Higgs mechanism \cite{SYmon,SYrev}.}
It is useful to introduce a gauge invariant order parameter
\beq
\Sigma={Y} \, X
\label{giop}
\eeq
which has the chiral transformation properties similar to
those of the ordinary
chiral condensate in QCD. The order parameter $\Sigma$
transforms nontrivially  under action the axial singlet $U(1)_A$ charge. The eight mesons
 parameterizing  the ``phases"  of the matrix $\Sigma$ are
\beq
\Sigma = |\Sigma| \,\exp\left (2i\, \frac{T^a \pi^a}{F_{\pi}}\right)
\label{dva}
\eeq
where $T^a$ are the SU(3) generators and $F_\pi$ is the ``pion" constant. As for the
absolute value of $\Sigma$,
\beq
|\Sigma| \propto \frac{\mu^4 \Delta_0^2}{g^2}
\label{tri}
\eeq
where $\Delta_0$ is the  superconducting gap at zero temperature
\beq
\Delta_0 \propto \mu \, \left(g(\mu )\right)^{-5}\, \exp\left({c}/{g(\mu )}
\label{chetyre}
\right),
\eeq
and $g$ is the QCD coupling constant.
Numerically $\mu$ is assumed to lie in the ballpark $$\mu\sim 1\,\,{\rm GeV}$$
 considered to be large in the scale $$\Lambda_{\rm QCD}\sim 200 \,\,{\rm MeV}\,.$$
The value of the gap parameter $\Delta_0$ is
$$
\Delta_0 \sim 10\,\, {\rm MeV}\,.
$$

The CFL mesons have to be considered as four-quark states
contrary to conventional two-quark Goldstone states in zero-density QCD.
However, let us emphasize that the Goldstone mesons in the CFL phase have
exactly the same
quantum numbers as the Goldstone mesons in QCD at at $\mu=0$.
The coupling constant $F_\pi$ in (\ref{dva}) is proportional  \cite{ss} to $\mu$, though.

The electromagnetic U(1)$_{Q}$ is broken by the condensates (\ref{odin}).
A closer look at these condensates shows,   however,
that a linear combination of the photon $A_{\mu}$ and $A_{\mu}^3,\, A_{\mu}^8$ gluons
\beq
\tilde{A}_\mu  \propto \left\{ A_\mu - \frac{e}{g}\left(A_\mu^3+\frac{1}{\sqrt 3}  A^8_\mu\right)\right\}
\label{pyat}
\eeq
remains massless (the corresponding charge is unbroken).
Instead of U(1)$_{Q}$, one can
consider the charges with respect to U(1)$_{\tilde{Q}}$.
In the CFL phase the mixing angle $\theta$ between the photon
and the  gluon component is obviously small.  Therefore, the massless gauge field is
is dominated by the original photon. Note that the off-diagonal gluons are charged
with respect to the U(1)$_{\tilde{Q}}$ charge.

\vspace{2mm}

When we switch on nonvanishing quark masses, strictly massless excitations disappear, of course.
A number of observables develop a rather contrived
mass hierarchy.  Our initial consideration will refer to a model which is semirealistic, at best.
However, it explains our basic observation in the most transparent way.
We will change some details later to make our model more realistic.

In the spirit of the Ginzburg--Landau theory of superconductivity
we will represent the diquark order parameters by a $3\times 3$ matrix of scalar fields $\Phi^{kA}$
where $k$ and $A$ are the color and flavor indices, respectively.

\section{Ginsburg--Landau effective description}
\label{gled}
\setcounter{equation}{0}

If the value of the  chemical potential $\mu$ is large, QCD is in the CFL phase. The
order parameter which develops a vacuum expectation value  (VEV) is the diquark condensate
\beq
\Phi^{kC} \sim  \varepsilon_{ijk}\,\varepsilon_{ABC}
\left( q^{iA}_\alpha\,  q^{jB\,\alpha}\, +
\bar{\tilde{q}}^{iA \,\dot{\alpha}}\,  \bar{\tilde{q}}^{jB}_{\dot{\alpha}}
\right),
\label{diquark}
\eeq
cf. Eq. (\ref{odin}),
where
$q_\alpha$ and $\tilde{q}_{\alpha}$ are left-handed spinors;
the first one is in the triplet representation
of the color and flavor SU(3) groups, while the second one is in the antitriplet representation.
The Dirac spinor of each
flavor is composed from a pair $(q_\alpha ,\, \bar{\tilde{q}}^{\dot\alpha})$.

At first, we consider dense QCD in the chiral limit, i.e. assume  that  quark masses  all vanish,
\beq
m_u=m_d=m_s=0\,.
\label{quarkmasses0}
\eeq
  Later (Section \ref{tamrs}) we will be able to relax this condition and switch on a nonvanishing
strange quark  mass. With the vanishing quark masses the symmetry group is
\beq
{\rm SU(3)}_C\times {\rm SU(3)}_{L}\times {\rm SU(3)}_R\times {\rm U}(1)_B
\label{symgroup}
\eeq
where SU(3)$_{L,R}$ are the global chiral groups associated with three flavors,
U(1)$_B$ is the global rotation associated with the
baryon number, while SU(3)$_{C}$ is the gauge group.

At small temperatures the gap (the diquark condensate)
is large, while the light degrees of freedom are the
Goldstone modes associated with the chiral rotations. These Goldstone modes
are quite similar to ordinary pions of zero-$\mu$ QCD.
The effective Lagrangian is a chiral Lagrangian for these
pions, see \cite{2} for a review.

At temperatures approaching (from below)   the critical temperature $T_c$,
 the gap becomes small, and its fluctuations become exceedingly more important.
It is assumed in what follows that the  chiral fluctuations are small (and can be ignored)
compared to the non-chiral gap fluctuations and and those of the gauge fields \footnote{Strictly speaking
at zero quark masses chiral fluctuations (''pions'') are massless and should be
included in the low energy effective theory. We discuss their impact in
 Sec. \ref{secmud}.}.
This regime can be described in terms of
a Ginsburg--Landau (GL) effective theory of  a complex matrix  scalar field $\Phi^{kA}$
defined in (\ref{diquark}) . The Ginsburg--Landau action has the form \cite{GLdesc,fthdqcd3}
\footnote{Here and below we use a formally
Euclidean notation, e.g.
$F_{\mu\nu}^2 = 2F_{0i}^2 + F_{ij}^2$,
$\, (\partial_\mu a)^2 = (\partial_0 a)^2 +(\partial_i a)^2$, etc.
This is appropriate since we are  going to study static (time-independent)
field configurations, and $A_0 =0$. Then the Euclidean action is
nothing but the energy functional.}
\beqn
S &=& \int {\rm d}^4x\left\{ \frac1{4g^2}
\left(F^{a}_{\mu\nu}\right)^{2}+ 3 \,{\rm Tr}\, ({\mathcal D}_0 \Phi)^\dagger \,({\mathcal D}_0 \Phi )
\right.
 \nonumber\\[3mm]
&+&
\left.
\,{\rm Tr}\, ({\mathcal D}_i \Phi)^\dagger \,({\mathcal D}_i \Phi )+
V(\Phi)\right\}
\label{glmodel}
\eeqn
with the potential
\beq
V(\Phi)=-m_0^2\, {\rm Tr}\,
\left(\Phi^\dagger \Phi\right) +
\lambda\, \left( \left[{\rm Tr}\,
\left(\Phi^\dagger \Phi\right)\right]^2+
{\rm Tr}\,\left[
\left(\Phi^\dagger \Phi\right)^2\right]\right),
\label{bulkpot}
\eeq
where
\beq
{\mathcal D}_\mu \, \Phi \equiv  \left( \partial_\mu
-i A^{a}_{\mu}\, T^a\right)\Phi\, ,
\label{dcde}
\eeq
and $T^a$ stands for the SU(3)$_{\rm gauge}$ generator,
while  $g^2$ is the QCD coupling constant.
The global flavor SU(3) transformations  are similar, with  $T^a$  acting on $\Phi$ from the right.
Various parameters in (\ref{bulkpot}) are defined as follows:
\beq
m_0^2 =\frac{48 \pi^2}{7\zeta(3)} T_c(T_c-T), \qquad
\lambda=\frac{18\pi^2}{7\zeta(3)} \frac{T_c^2}{ N(\mu)}\,,
\label{stuff}
\eeq
while $N(\mu)=\mu^2/(2\pi^2)$ is the density of states on the Fermi surface. Note that
our field $\Phi$ in (\ref{glmodel}) is
 rescaled  as compared to that in~\cite{GLdesc,fthdqcd3}:  its kinetic energy is canonically normalized.

The critical temperature $T_c$ is
much smaller than $\mu$,
\beq
T_c \sim \frac{\mu}{\left(g(\mu )\right)^5}\, \exp{\left(-\frac{3\pi^2}{\sqrt{2}g(\mu )}\right)}\ll \mu\,.
\label{gap0}
\eeq
implying
\beq
m_0^2 \sim T_c(T_c-T), \qquad
\lambda\sim \frac{T_c^2}{\mu^2}\ll 1,
\label{stuffestimates}
\eeq
Minimizing  (\ref{bulkpot}) we find
the $\Phi$ field VEV,
\beq
\Phi_{\rm vac} = v\,{\rm diag}\, \{1,1,1\}\,,
\label{diagphi}
\eeq
where the value of the parameter
 $v$ is given by\,\footnote{Due to the $\Phi$ field rescaling in (\ref{glmodel})  our VEV $v\sim \mu$
is different from the standard definition of the gap which is
$\Delta \sim T_c $.}
\beq
v^2 = \frac{m_0^2}{8\lambda}=\frac{4\pi^2 }{3}
\,\frac{T_c-T}{T_c}\,\mu^2\,.
\label{gap}
\eeq

The diagonal form (\ref{diagphi}) of the vacuum $\Phi$ matrix,  of the Bardakci--Halpern type
\cite{BarH}, expresses the phenomenon of the
color-flavor locking.
The gauge symmetry of the Lagrangian is SU(3)
while its flavor symmetry is SU(3)$\times$U(1), see (\ref{symgroup}).
The symmetry of the vacuum state is the diagonal SU(3)$_{\rm CF}$,
\beq
{\rm SU(3)}_C\times {\rm SU(3)}_{F}\times {\rm U}(1)_B \to
{\rm SU(3)}_{\rm CF} \,.
\label{symbreak}
\eeq
Nine symmetries are spontaneously broken.
Out of nine Goldstone modes in this model,
eight are eaten up by the gauge bosons, which are fully Higgsed, while
one -- a common phase of the matrix $\Phi$ -- survives as a massless excitation. It is associated with broken global baryon symmetry U(1)$_B$.

The spectrum of massive excitations around this vacuum
can be read off  \cite{fthdqcd3} from the GL model Lagrangian (\ref{glmodel}).
The Higgsed gluons acquire masses
\beq
m_g =g v\sim g\mu \, \sqrt{\frac{T_c-T}{T_c}}\,,
\label{gmass}
\eeq
while nine remaining scalars of the complex matrix $\Phi^{kA}$
fill singlet and octet representations of the unbroken
SU(3)$_{\rm CF}$. Their masses are
\beq
m_1 = 2m_8 = \sqrt{2}\,m_0 \sim \sqrt{T_c(T_c-T)}.
\label{smass}
\eeq
Since $m_g \gg m_0$, as a consequence of the condition
$T_c\ll \mu$,
 we   deal here with the extreme type I
superconductivity \cite{GiannRen}.

The GL model (\ref{glmodel}) is in the weak coupling regime if
we assume that
\beq
m_g\gg \Lambda_{\rm QCD}.
\label{wccondition}
\eeq

\section{Non-Abelian strings in the CFL phase}
\label{nasicflp}
\setcounter{equation}{0}

The model (\ref{glmodel}) we  use for our analysis is similar to that
 in which the non-Abelian strings were first
considered in the non-supersymmetric setting \cite{gsy}.
Compared to the original version \cite{gsy}, we
discard the U(1) gauge bosons, since in high-density QCD
only  the non-Abelian color SU(3)   is gauged.
The baryon current is not gauged, while the photon interaction with the
electromagnetic current can be neglected for the time being due to its weakness
compared to the quark-gluon interactions (see, however, Section \ref{tamrs}).
This means that the vortices we will deal with are not fully local. In
their U(1) part they are global.
This would make their tension infinite if the perpendicular dimensions were infinite
too.
In the context of dense QCD, with finite-size samples, this is not a problem:
the logarithmic divergence of the tension will be cut-off by the sample size.

The existence of non-Abelian strings, with the tension three times smaller than that of the U(1) global
string, is due to the fact that the $Z_3$ center elements of SU(3)$_{\rm gauge }$
simultaneously belong to the global U(1).
Everybody knows that $\pi_1({\rm SU}(3))$ is trivial.
In arranging a topologically nontrivial winding of the
scalar fields on the large circle encompassing the $Z_3$ string axis one can do the following.
The needed $2\pi$ winding will be split in two parts (there are three possible options):
$\pm 2\pi / 3$ will come from U(1) while the remainder will come from
rotations in the Cartan subgroup of SU(3) (in other words,
rotations around the third and the eighth axes, with the generators $T^{3,8}$), for instance,
\beqn
\Phi (r\to \infty ,\alpha ) &=& v\,{\rm diag } \,(e^{i\alpha}, 1 , 1)\,,
\nonumber\\[3mm]
A_i (r\to \infty ) &=& \frac{\varepsilon_{ij}x^{j}}{ r^2} \,{\rm diag}\,\left(-\frac{2}{3}, \frac{1}{3}, \frac{1}{3}\right)
\,,
\label{stringas}
\eeqn
where $i,j=1,2 $ are the directions perpendicular to the string axis
and $\alpha$ is the polar angle in the $12$ plane.
The topological stability of the straight $Z_3$ strings
is due to the fact that
$$
\pi_1\left (\frac{{\rm SU}(3)_{\rm CF}\times {\rm U}(1)_{\rm F}}{(Z_3)_{\rm CF}}\right)
$$
is nontrivial. There are three distinct $Z_3$ strings. Assembling all three such strings in one straight line
one obtains a string with triple tension topologically equivalent to the global U(1) string.

It is rather obvious that on each of the $Z_3$ strings
the diagonal SU(3) symmetry is further broken down to
SU(2)$\times$U(1). Now, one can construct
non-Abelian strings out of the $Z_3$ strings.
To this end one must rotate the given $Z_3$ solution
 inside the unbroken diagonal ${\rm SU}(3)_{\rm CF}$.
 This costs no energy; therefore orientational moduli
 associated with these rotations appear. Since the symmetry breaking pattern is
 $$
 {\rm SU}(3)_{\rm CF} \to  {\rm SU}(2) \times {\rm U}(1)\,,
 $$
 one has four moduli fields on the string world sheet,
 with the CP(2) target space. We refer the reader to the reviews \cite{Trev,Jrev,SYrev,Trev2}  for a detailed discussion.

 The string solution
involves the  global ${\rm U}(1)$;  hence, it contains a power tail
from the uneaten Goldstone boson. This tail results in the
logarithmic divergence of the string tension which is well familiar
to the global strings explorers. We have already mentioned this circumstance above.

More specifically, following the general procedure (see \cite{gsy}
or the
review paper \cite{SYrev}) we parametrize the solution for one of the $Z_3$ strings, say, that
 in (\ref{stringas}), as follows:
\beqn
\Phi (r ,\alpha ) &=& {\rm diag } \,(e^{i\alpha}\phi_1, \phi_2 , \phi_2)\,,
\nonumber\\[3mm]
A_i (r ) &=& \frac{\varepsilon_{ij}x^{j}}{ r^2} \,(1-f)\,{\rm diag}\,\left(-\frac{2}{3}, \frac{1}{3}, \frac{1}{3}\right),
\label{z3string}
\eeqn
where $\phi_1(r)$, $\phi_2(r)$,  and $f(r)$ are scalar and gauge
profile functions of the string, respectively. They satisfy the obvious  boundary conditions
\beqn
&& \phi_{1}(0)=0,
\nonumber\\[2mm]
&& f(0)=1,
\label{bc0}
\eeqn
at $r=0$, and
\beqn
&& \phi_{1}(\infty)=v,\;\;\;\phi_2(\infty)=v\,,
\nonumber\\[2mm]
&& f(\infty)=0,
\label{bcinfty}
\eeqn
at $r=\infty$.
Then the solution for the non-Abelian string can be written
as \cite{gsy,SYrev}
\beqn
\Phi &=& e^{i\alpha/3}\frac13\,[2\phi_2 +\phi_1] +e^{i\alpha/3}(\phi_1-\phi_2)\left(
n\,\cdot \bar{n}-\frac13\right) ,
\nonumber\\[3mm]
A_i &=& \left( n\,\cdot \bar{n}-\frac{1}{3}\right)
\varepsilon_{ij}\, \frac{x_j}{r^2}
\,
f(r) \,,
\label{nastr}
\eeqn
where $n^A$ ($A=1,2,3$) are complex orientational moduli\,\footnote{$A_i $ is a matrix;
correspondingly, $ n\,\cdot \bar{n}$ should be understood as
 $n^i\,  \bar{n}_j.$ The bar stands for the complex conjugation.}
 of the
string parameterizing the CP(2) moduli space, $|n^A|^2=1$.
The profile functions of the non-Abelian strings satisfy the second order
equations of motion,
\beqn
&&
f'' -\frac{f'}{r} -\frac{g^2}{3}\,(1+2f)\,\phi_1^2
+\frac{g^2}{3}\,(1-f)\,\phi_2^2=0,
\nonumber\\[3mm]
&&
\phi_1''+\frac{\phi_1'}{r} - \frac1{9}\,\frac{(1+2f)^2}{r^2}\,
\phi_1 -\frac12 \frac{\pt V}{\pt \phi_1} =0,
\nonumber\\[3mm]
&&
\phi_2''+\frac{\phi_2'}{r} - \frac1{9}\,\frac{(1-f)^2}{r^2}\,
\phi_2 -\frac14 \frac{\pt V}{\pt \phi_2} =0,
\label{soeqs}
\eeqn
where primes denote differentiations with respect to $r$.
Here $V$ is given by (\ref{bulkpot}), and we note that
$\pt V/\pt \phi_{1,2} \sim m_0^2 (\phi_{1,2}-v)$ are rather small
due to the smallness of $m_0$.

These equations were studied  in \cite{EtoNitta}.
A feature of the non-Abelian string in the CFL phase is the
 power fall-off of the singlet scalar profile function at spatial
infinity due to the presence of the corresponding massless Goldstone.
From (\ref{soeqs}) we find that at $r\to\infty$ ($r\gg1/m_0$) the
profile functions behave as \cite{EtoNitta}
\beqn
&&
\phi_1 \sim \phi_2\sim v\left( 1- \frac1{3m_1^2r^2}+\dots\right),
\nonumber\\[3mm]
&&
(\phi_1-\phi_2)\sim v\,e^{-m_0 r}, \qquad f\sim e^{-m_0 r}.
\label{profileinfty}
\eeqn
In \cite{EtoNitta}
solutions to Eq. (\ref{soeqs}) were found numerically.
Here, to study these equations {\em analytically}, we will apply the method
used for the Abelian strings in the extreme type I superconductors \cite{Y99}.
At distances $r\lsim 1/m_0$, the string profile functions in fact have
a two-scale structure due to smallness of the  ratio of the scalar to gluon mass.
 At   $1/m_g\ll r \ll 1/m_0$,
 the  gluons can be considered as heavy, and we can neglect the gauge  kinetic term.
 This boils down to dropping two first terms in the first equation in (\ref{soeqs}). Then,
  this equation becomes algebraic and yields
\beq
f\approx\frac{\phi_2^2-\phi_1^2}{\phi_2^2+2\phi_1^2}\,.
\label{f}
\eeq
Substituting this result in the two last equations in (\ref{soeqs}) we
find  the approximate solution    in a simple
form
\beqn
&&
\phi_1 \approx b\,v\,(m_0 r) +\dots,
\nonumber\\[3mm]
&&
\phi_2 \approx v\left[1 +O((m_0 r)^{4})\right],
\nonumber\\[3mm]
&&
f\approx 1-3b^2\,(m_0 r)^{2}+\dots ,
\nonumber\\[3mm]
&&
1/m_g\ll r \ll 1/m_0\,,
\label{profileinter}
\eeqn
where the expansion goes in powers of $(m_0 r)^2$,
while $b$ is a number, $b\sim 1$.
Here we use the fact
that both the singlet and octet scalar masses
are of the same order $\sim m_0$ much less then the gluon mass $m_g$, see (\ref{gmass}), (\ref{smass}).

Properties of the $Z_3$ strings in application to the CFL phase of dense QCD
were also considered in  \cite{mat, gm,fthdqcd1,iida,forbes}.

\section{On the string world sheet}
\label{osws}
\setcounter{equation}{0}

The low-energy description of massless excitations on the string world sheet
includes two decoupled sectors: the two translational moduli and four orientational.
The translational moduli are described by the Nambu--Goto action with the constant $T$
which logarithmically diverges,
\beq
S_{\rm NG} =T_0 \int d^4 x\,\,{\mathcal{L}}_{\rm NG}  \,,\qquad T_0 = 2\pi\, v^2 \, \ln\left(L m_0 \right)\,,
\label{ws}
\eeq
where $L$ is a typical size of the color-flavor locked medium. This part
is well known and will be of no concern to us here.

The orientational moduli's interaction
is governed by  CP($N-1$) (with $N=3$ in the case at hand). In the non-supersymmetric setting this
model was shown to appear on the world sheet of a non-Abelian string in \cite{gsy}. For non-Abelian strings in the model (\ref{glmodel}) the effective world-sheet theory was obtained in
\cite{nittaws}. In the gauged formulation the CP(2)
model  takes the form
\beq
S_{\rm CP(2)}   = 2\beta  \,\int d t\,d x_3 \left\{
3\left| {\mathcal D}_0 n^A\right|^2 +
\left| {\mathcal D}_3 n^A\right|^2 \right\}\,,
\label{sac}
\eeq
where  $\beta$ is the CP(2) coupling constant, while
the complex fields $n^A$  ($A=1,2, 3$)
are orientational moduli of the string promoted to world-sheet fields , see (\ref{nastr}). They
 transform in the fundamental representation of
SU(3).
These fields are subject to the constraint
\beq
\bar n_A \, n^A =1\,.
\label{consed}
\eeq
 The $n$ fields have a U(1) charge which is gauged,
\beq
{\mathcal D}_\alpha n^A \equiv \left(\partial_\alpha - i A_\alpha\right) n^A\,.
\eeq
The two-dimensional photon field $A_\mu$ has no kinetic term in the
Lagrangian (\ref{sac}) and can be viewed as auxiliary,
\beq
A_{\alpha} =\frac{i}{2}\left( \bar n_A \stackrel{\leftrightarrow}{\partial_\alpha} n^A\right)\,.
\label{amuex}
\eeq
It does acquire a kinetic term in the solution of the model, however,
which plays an important dynamical role.

The coupling constant $\beta$ is determined by substituting
the solution for the non-Abelian string (\ref{nastr}) in the kinetic terms of the bulk action (\ref{glmodel}) and assuming that the
moduli $n^A$ has a slow adiabatic dependence on the the world sheet coordinates $t$ and $x_3$. We
also use  the following  expressions for the
$A_0$ and $A_3$ components of the gauge potential \cite{gsy,SYrev}:
\beq
A_{\alpha}=-i\,  \big[ \pt_{\alpha} n \,\cdot \bar{n} -n\,\cdot
\pt_{\alpha} \bar{n} -2n\,
\cdot \bar{n}(\bar{n}\pt_{\alpha} n)
\big] \,\rho (r)\, , \quad \alpha=0, 3\,,
\label{An}
\eeq
where we  introduce a new profile function $\rho (r)$
with the boundary conditions
\beq
\rho (\infty)=0, \qquad \rho (0)=1\,.
\label{bcrho}
\eeq
The function $\rho (r)$ in Eq.~(\ref{An}) is
determined  through a minimization procedure  \cite{ABEKY,SYmon,gsy,SYrev}
which generates $\rho$'s own equation of motion.

This procedure leads us to the CP(2) model (\ref{sac}) where
coupling constant $\beta$ is determined by the integral
\beqn
\beta & = &
  \frac{2\pi}{g^2}\,\int_0^{\infty}
rdr\left\{\left(\frac{d}{dr}\rho (r)\right)^2
+\frac{1}{r^2}\, f^2\,\left(1-\rho \right)^2
\right.
\nonumber\\[4mm]
& + &
\left.  g^2\left[\frac{\rho^2}{2}\left(\phi_1^2
 +\phi_2^2\right)+
\left(1-\rho \right)\left(\phi_2-\phi_1\right)^2\right]\right\}\, .
\label{I}
\eeqn
Then the equation for the profile function $\rho$ is
\beq
-\frac{d^2}{dr^2}\, \rho -\frac1r\, \frac{d}{dr}\, \rho
-\frac{1}{r^2}\, f^2 \left(1-\rho\right)
+
\frac{g^2}{2}\left(\phi_1^2+\phi_2^2\right)
\rho
-\frac{g^2}{2}\left(\phi_1-\phi_2\right)^2=0\, .
\label{rhoeq}
\eeq

In Ref.~\cite{nittaws} it was shown that,
despite of the presence of the power fall-off of the string solution,
 the coupling $\beta$ is finite, i.e. the orientational modes of  non-Abelian string are normalizable. The reason for this is that the only profile function with the power fall-off is that of
the singlet component of the scalar field \cite{EtoNitta},
see also (\ref{profileinfty}).
On the other hand, the coupling $\beta$ is associated with dynamics of the string orientational moduli and, therefore,  is determined by the octet  component of the scalar field and the gauge
profile functions which have the exponential fall-off. In particular, at $r\to\infty$,
the  function $\rho$ behaves as
$$\rho\sim e^{-2m_0 r}\,.$$

The main contribution to the integral in (\ref{I}) comes from
the region of intermediate $r$,
$$1/m_g\ll r \lsim 1/m_0\,.$$ In this domain we  can
neglect, as previously, the kinetic term of the gauge field. This leaves us with only two last terms in
Eq. (\ref{rhoeq}). The equation becomes algebraic, and  we  can can write
\beq
\rho\approx \frac{(\phi_2-\phi_1)^2}{\phi_1^2+\phi_2^2}\,.
\label{rhosol}
\eeq
Substituting here the expansions (\ref{profileinfty}) and (\ref{profileinter})  we get an
estimate for the coupling $\beta$,
\beq
\beta \approx \pi\int_{0}^{\infty} rdr \frac{(\phi_2^2-\phi_1^2)^2}{\phi_1^2+\phi_2^2}
=  c\,\frac{v^2}{m_0^2} \sim \frac{\mu^2}{T_c^2}\gg 1\,.
\label{beta}
\eeq
Unfortunately, we cannot calculate the constant $c$ from
the expansions (\ref{profileinfty}) and (\ref{profileinter}); we only know that $c\sim 1$.
Numerically, $\beta$ was calculated in \cite{nittaws} for different
values of the masses $m_g$, $m_1$ and $m_8$. On the other hand, our estimate has all virtues of
the analytic expression.

We see that $\beta$ is rather large. The reason for this is that the
fall-off of the string solution is controlled by the small scalar mass $m_0$.

In quantum theory the coupling constant of the CP(2) model runs. The
CP($N-1)$ models are asymptotically free and generate their own scale $\Lambda_{CP}$.
The estimate (\ref{beta}) is classical and  refers to the scale which determines the inverse thickness of the string \cite{SYrev} given by $m_0$. This is because the CP(2) model (\ref{sac}) is an effective low-energy theory on the string world sheet. Its physical ultraviolet cut-off is given by the inverse thickness of the string. This implies
\beq
4\pi\beta(m_0) = N \,\ln{\frac{m_0}{\Lambda_{CP}}}\,, \qquad N=3\,,
\label{running}
\eeq
an equation determining the scale $\Lambda_{CP}$ of the effective world sheet theory for the non-Abelian string. From (\ref{running}) we get
\beq
\Lambda_{CP}= m_0\,\exp{\left(-\frac{4\pi c }{N} \,\frac{v^2}{m_0^2}\right)}\ll m_0, \qquad N=3.
\label{lambdaCP}
\eeq
We see that $\Lambda_{CP}$ is exponentially
 small. Note that for the BPS-saturated strings
in \ntwo supersymmetric QCD the relevant parameter $\Lambda_{CP}$ turns out to be equal
to the scale $\Lambda$ of the bulk theory. This feature is specific for \ntwo supersymmetry.
In the  CFL phase of dense QCD such an equality  does not hold. The reason is that  we
deal with the extreme type I superconductivity in the case at hand.

\section{Kink-antikink mesons at large $N$}
\label{kkmesons}
\setcounter{equation}{0}

The CP($N-1$) model at large $N$ was solved in \cite{AddaS,WitS}
and the qualitative features of this solution are known to stay valid down to
$N=3$ and even $N=2$. Below we will outline the features
which are important for our purposes.

\begin{figure}[h]
\centerline{\includegraphics[width=2.5in]{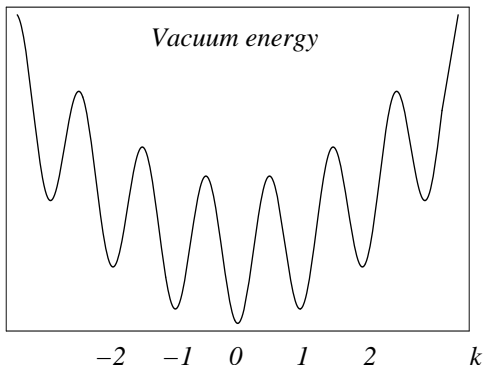}}
\caption{\small
The vacuum structure of the
CP$(N-1)$ model with the
vanishing vacuum angle. The genuine vacuum is labeled by $k=0$.
All minima with $k\neq 0$ are quasivacua.}
\label{stablequasi}
\end{figure}

In the  CP$(N-1)$ model
the genuine vacuum state is unique. However,
 there are of order $N$ quasivacua \cite{wittenproved} (local minima of the ``potential"), which lie
higher in energy than the genuine one.
(Fig.~\ref{stablequasi}).

 A family of quasi-vacua
 have  energies
\beq
 E_k  \sim \, N\, \Lambda_{CP}^2
\left\{1 + {\rm const}\left(\frac{2\pi k}{N}
\right)^2
\right\}
\,,\,\,\,\, k=0\ldots, N-1\,.
\label{split}
\eeq
The energy split between two neighboring (quasi)vacua is $O(1/N)$.
In fact, the fields $n$ and $\bar n$
represent kinks and antikinks interpolating between the genuine vacuum and its neighbors.
Since their energies are non-degenerate neither kinks nor antikinks
can exist on the string in isolation. Only the kink-antikink pairs -- mesons -- are allowed.

\begin{figure}[h]
\centerline{\includegraphics[width=3.5in]{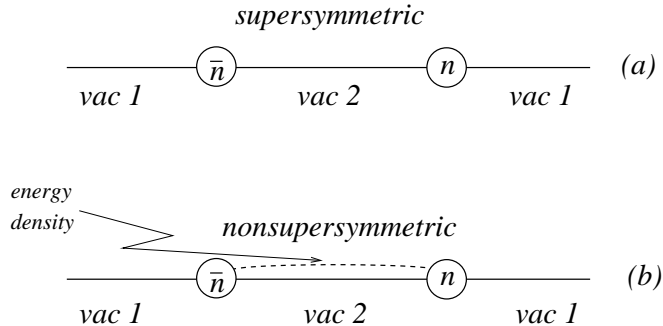}}
\caption{\small
A kink-antikink state in the   CP($N-1$) model. In the nonsupersymmetric version the
vacuum 2 is a quasivacuum whose energy density is   higher
than that of the genuine vacuum 1 by $\sim \Lambda_{CP}/N$.}
\label{kinkantikin}
\end{figure}

A kink-antikink configuration on the flux tube is
shown in Fig.~\ref{kinkantikin}.
 It is pretty obvious that the energy of this configuration linearly depends on
the distance between $n$ and $\bar n$, so that these kinks are confined along the string and form a meson.

We see that the splitting at large $N$ is $\Lambda_{CP}^2/N$, while the mass of an individual
kink is of the order of $\Lambda_{CP}$,
\beq
m_{\rm kink}\sim \Lambda_{CP}\,.
\label{kinkmass}
\eeq
Thus, the distance between
kinks in the meson is $\sim N/\Lambda_{CP}$, i.e. much larger than the size
of the individual kink ($\sim 1/\Lambda_{CP}$). Hence, the kink and
antikink are well separated inside the bound state, the meson.
Since the $n$ kinks are in the fundamental
representation of the global SU$(N)_{\rm CF}$ symmetry \cite{WitS,HoVa}, the mesons can be of two types:
SU$(N)_{\rm CF}$ singlets or  adjoints. Mesons with the  quantum numbers
of the adjoint representation cannot decay  because the
global SU$(N)_{CF}$ is unbroken and this global quantum number is conserved. Given that kink and antikink are well separated inside such a meson (provided the string is sufficiently long, which we assume, of course)
it is clear that the kink-antikink
mesons are in fact the lightest adjoint states and are {\em stable}.
This conclusion is supported by the exact solution of the CP(1) model \cite{zamo}
demonstrating stability of the kink-antikink mesons even at $N=2$.

\section{Confined monopoles in the CFL phase}
\label{moncfl}
\setcounter{equation}{0}

What is the bulk interpretation of kinks interpolating between
the neighboring (quasi)vacua of world-sheet CP$(N-1)$ model?
This problem was studied in detail in supersymmetric gauge theories \cite{HT1,SYmon,HT2,SYrev},
and the answer is known.

The non-Abelian strings were first found in \ntwo supersymmetric
 QCD with the gauge group U$(N)$ and $N_f=N$ where $N_f$ is the number of the quark flavors.
The scalar quarks (squarks) develop condensate of the color-flavor locked form which
  leads to   formation of the non-Abelian strings. \ntwo supersymmetric
 QCD have adjoint scalar fields along with the gauge fields form the \ntwo vector supermultiplet. These adjoint scalars develop  VEVs as well, Higgsing
 the gauge U$(N)$ group
 down to its maximal Abelian subgroup,
 which ensures existence of the conventional  't Hooft-Polyakov monopoles in the theory.
 The squark condensates then break the gauge U$(N)$ group completely, Higgsing all gauge bosons.
Since  the gauge group is fully Higgsed
the 't Hooft-Polyakov monopoles
are {\em confined}. As we know, in the Higgsed U$(N)$ gauge theories
the magnetic
monopoles show up
only as junctions of two distinct elementary non-Abelian strings \cite{Tong,SYmon,HT2}.

Now, let us verify that the
confined magnetic monopole in the case at hand, CFL phase of dense QCD,
is a junction of two strings seen in the two-dimensional
$CP(N-1)$ model at
 $N=3$.
Consider the junction of two $Z_3$ strings given by (\ref{nastr}).
Three distinct $Z_3$ strings
mentioned above
correspond to three choices of the orientation vector
$n^A$,
$$n^A=(1,0,0)\,,\quad n^A=(0,1,0)\,,\,\,\,\mbox{and}  \,\,\, n^A=(0,0,1)\,,$$ see (\ref{z3string}).
The magnetic flux of the junction of say, $n^A=(1,0,0)$ and  $n^A=(0,1,0)$
strings  is
given by the difference of the fluxes of these two strings. Using (\ref{nastr})
we get that the flux emanating from  the junction is
\beq
  4\pi\,\times \, {\rm diag} \, \frac12\, \left\{ \,1 ,\,  -1,\,0 \right\} \,,
\label{monflux}
\eeq
This is exactly the  flux of the 't Hooft--Polyakov
monopole with the magnetic charge
\beq
(n_3,n_8)= (1,0)\,,
\label{12mon}
\eeq
where $(n_3,n_8)$ are the  magnetic charges with respect to
$T_3$ and $T_8$ generators of SU(3)$_{\rm gauge }$, respectively.

Similarly, two other string junctions, namely     $n^A=(0,1,0)$, $n^A=(0,0,1)$
and $n^A=(0,0,1)$, $n^A=(1,0,0)$, have the magnetic fluxes equal to the
fluxes of two other elementary monopoles in SU(3) (i.e. given by
 two other roots of SU(3) algebra), namely,
\beq
(n_3,n_8)= \left(-\frac12,\frac{\sqrt{3}}{2}\right), \qquad (n_3,n_8)\left(-\frac12,-\frac{\sqrt{3}}{2}\right).
\label{othermon}
\eeq

How this picture is seen in the effective world-sheet theory on
the non-Abelian string? For \ntwo supersymmetric bulk theory
the world-sheet  CP$(N-1)$ model is also supersymmetric and has $N$ {\em degenerate}
 vacua. The
elementary non-Abelian strings  are in fact represented by $N$ distinct vacuum states in the effective
world-sheet  CP$(N-1)$ model,   while the
 confined monopoles are  the kinks interpolating
between these distinct vacua \cite{SYmon,HT2,Tong}.

Now, we can add a mass $m_{\rm adjoint}$ to the adjoint scalars,
increase it eventually tending to infinity and decoupling these adjoint scalars from the bulk theory
(this breaks supersymmetry in the bulk down to \none ). What happens to the confined monopoles?
The orientational moduli
of the non-Abelian fluctuate, with their fluctuations becoming
exceedingly stronger as  $m_{\rm adjoint}$ grows. When $m_{\rm adjoint}\gg \Lambda$
strings' fluxes  no longer particular directions. Fluxes are  smeared over the whole
group space. Since the adjoint scalars are no longer present in the theory,
 naively it seems that there are no 't Hooft-Polyakov monopoles.
 At least they are not seen in the quasiclassical approximation.

In the world-sheet theory this corresponds to a highly quantum,
strong coupling regime. However, we know that in this regime
still there are $N$ degenerate vacua in the world sheet CP$(N-1)$
model \cite{mmodel,SYhetN,SYrev}. Moreover, there are kinks interpolating
between these vacua. They are stabilized by quantum effects and have a
non-zero mass (of order $\Lambda_{CP}$)  and finite size
(of the order of $\Lambda_{CP}^{-1}$). It is clear that these kinks
correspond to confined monopoles of the bulk theory
in the highly quantum regime. More exactly, we should say that they
represent   what {\em becomes} of the confined 't Hooft--Polyakov monopoles
deep in the quantum non-Abelian regime when they cannot be seen
in the quasiclassical approximation.

If we break supersymmetry in the bulk even further, down to nothing,
 the vacuum
energies of the CP($N-1$) model split, as was explained in the previous section \cite{gsy}.
Kinks and antikinks are confined along the string and form
kink-antikink mesons. The
kink confinement in the two-dimensional CP model can be interpreted \cite{gsy} as
the following phenomenon:
the non-Abelian monopoles,
in addition to  the four-dimensional confinement
(which ensures that the monopoles
are attached to the strings) acquire a two-dimensional confinement along
the string:
a monopole--antimo\-nopole  pair forms a meson-like configuration,
with necessity.

Moreover,  as was shown in \cite{WitS,HoVa} for the \cpn model,
the kinks belong to the fundamental
 representation of the global group SU$(N)_{\rm CF}$.
 Therefore,
the monopole-antimonopole mesons which belong to the adjoint representation
 with respect to this group
are stable.

What lessons can  we  learn from this ``supersymmetry saga''
for dense QCD?
We might think that there is a certain deformation of the GL model
(\ref{glmodel}) which includes adjoint scalar fields.
If these fields develop VEVs, the conventional  't Hooft-Polyakov monopoles are formed.
If in addition, the diquark condensate that develops in the color-flavor
locked phase, produces non-Abelian strings
which confine these monopoles attaching them to the strings.
These confined monopoles are seen as loosely bound kinks in the CP$(N-1)$ model
on the string world sheet.

Now we give a mass to the adjoint scalars, eventually decouple them in the bulk
theory ending up with our GL model (\ref{glmodel}). No monopoles can  now be found in
the quasiclassical approximation. However, we are aware of their presence: they manifest themselves
as
confined monopoles   seen as kinks on the non-Abelian strings.

Due to the quantum splitting of the string tensions (the vacuum energies in CP$(N-1)$ model
split, see (\ref{split})) the magnetic monopoles cannot
move freely along the string. Monopoles are bound with antimonopoles in the monopole-antimonopole mesons on the string.
However, as long as the string is very long
and the splitting is small, the monopole and antimonopole in the pair
are well separated inside the meson, and  our conclusion of the presence of
the confined monopoles in the CFL phase of QCD stays intact.

It is plausible to suggest that these monopoles become unconfined
as we reduce $\mu$ and cross the phase transition line into
the normal phase of QCD at smaller temperatures and chemical potentials.
Although this is a pure speculation, of course, but, if so,
they might  condense in this phase triggering the quark confinement.

\section{Towards a more realistic setting}
\label{tamrs}
\setcounter{equation}{0}

Now we will try to work out a more realistic setup, to make
details of our analysis   closer related to   actual
  dense QCD, with three flavors of unequal mass.

\subsection{$N=3$}
\label{sub81}

In the real world $N$ is not so large,   $N=3$.
Therefore, strictly speaking, the large-$N$ solution of
the CP$(N-1)$ model \cite{WitS} briefly reviewed in Section \ref{kkmesons}
does not apply.
However, as was mentioned in Section \ref{kkmesons},
the qualitative features of this solution are valid for not-so-large $N$ such as $N=3$ and
even $N=2$.  Indeed, at $N=2$ the model was solved exactly \cite{zamo}.
Zamolodchikovs found that the spectrum of the
O(3) model (equivalent to the CP(1) model) consists of a triplet of degenerate states
(with mass $\sim \Lambda_{CP}$).
At $N=2$ the action (\ref{sac}) is built of doublets.
In this sense one can
say that Zamolodchikovs' solution exhibits confinement of
doublets. This is in qualitative accord with the large-$N$ solution \cite{WitS}.

In our case ($N=3$) we have two quasivacua, in addition to the lowest-energy state, the vacuum,
 and a triplet of kinks which,
because of their linear attraction, are confined and form  singlet and octet mesons on the string.

The question  we address here is: given a state belonging
to the adjoint representation of SU(3)$_{CF}$, can we say whether this state is a
``perturbative" Higgsed gauge boson, or
a kink-antikink meson? In other words, if the kink comes too close to antikink,
  can they annihilate into the perturbative state with  the same
SU(3) global charge?

Here the analogy with the supersymmetric setting is useful (although the string
vacua are not split
 in supersymmetric CP$(N-1)$ models). In the quasiclassical
 regime, outside the so-called curves of the marginal stability
(CMS),  perturbative  states are present in the spectrum of the
CP($(N-1)$ model, while inside CMS, in the strong coupling domain, they just
do not exist as stable states. They
decay into the kink-antikink  pairs \cite{Dorey,HaHo,SYtorkink}.
In the non-supersymmetric CP($N-1$) models we do not have that degree of
 control over the spectrum, while the curves of the marginal stability
 are replaced by the phase transition lines.
However, qualitatively, the analogy with supersymmetric case is instructive
Due to
string vacua splitting the  kink and antikink cannot fly apart along the string,
If $N$ is large we have well-separated monopole and
antimonopole in such a meson, as was explained in Sec. \ref{kkmesons}.
Even if $N$ is not large, say, $N=2$, it is almost impossible to
think that the Zamolodchikov
triplet is anything other than
 a kink-antikink bound state.
This is in line with the supersymmetric case which shows that there are no
``perturbative"  stable states inside CMS in the  strong coupling domain.

Now, let us   translate this two-dimensional picture in terms
of strings and mono\-poles of the bulk theory. At large
$N$ monopoles and antimonopoles are well separated on the
non-Abelian string at hand,
and we can clearly  identify  these states in the theory. Moreover, even if
$N$ is not that large, the analogy with the supersymmetric set-up
tells us that the adjoint string-attached mesons
 are most likely formed by monopole-antimonopole pairs.
 This is true as long as the splitting of non-Abelian strings is a quantum strong
 coupling effect determined by the CP$(N-1)$ scale $\Lambda_{CP}$, see (\ref{split}).

\subsection{Non-zero strange quark mass}

Now we will move on towards reality in another direction,
considering  the effect of   the nonvanishing strange quark mass
$m_s$ (we will continue to assume that the $u$ and  $d$-quarks are strictly massless),
\beq
m_u=m_d=0,\qquad m_s\neq 0.
\label{nonzeroms}
\eeq
For small nonvanishing $m_s$ the GL potential (\ref{bulkpot})
acquires the following correction \cite{Iidams}:
\beq
\delta V(\Phi)= \epsilon \, \left\{
\Phi^{\dagger}_{u} \Phi^{u}
+\Phi^{\dagger}_{d} \Phi^{d}\right\},
\label{potcorr}
\eeq
which affects only quadratic terms in the Lagrangian, while the
corrections to $\lambda$ is negligible. Here the contraction of color indices is assumed, while
the parameter $\epsilon$ is
\beq
\epsilon=\frac{48 \pi^2}{7\zeta(3)} \,\frac{m_s^2}{4\mu^2}\,T_c^2\,\ln{\frac{\mu}{T_c}}\,.
\label{epsilon}
\eeq
As a result, the VEVs of the diquark fields change, namely,
\beq
\langle\Phi \rangle ={\rm diag}\left( v_u,v_u,v_s\right)\,,
\label{qvevdef}
\eeq
where
\beq
v_u^2= \frac{m_0^2- 2\epsilon}{8\lambda}, \qquad
v_s^2= \frac{m_0^2+2\epsilon}{8\lambda}\,.
\label{vuvs}
\eeq
We see that the $us$ and $ds$ condensates are smaller than the $ud$ condensate.
The nonvanishing difference between $v_s$ and $v_u=v_d$ breaks the residual global color-flavor
symmetry,
\beq
{\rm SU(3)}_{ CF}\to {\rm SU(2)}_{ CF}\times {\rm U}(1)\,.
\label{newc+f}
\eeq
Only the reduced color-flavor locked  SU(2)$_{ CF}$
survived the perturbation.

It is worth mentioning that when we switch on
the strange quark mass  perturbation in (\ref{potcorr}), we do not take into account the
effect of the overall charge
charge neutrality.
The latter leads to  shifts in the chemical potentials of $u$, $d$ and $s$ quarks proportional
to the chemical potential of the electrons \cite{Iidams}. These shifts effectively further break
SU(2)$_{ CF}$ because the electric charges of the $u$ and $d$-quarks
are different.
For our purposes it is alright to neglect this effect since we
 do not address  in this paper dynamical processes in the actual neutron stars.

Instead, we consider a {\em gedanken} QCD at large chemical potential $\mu$,
switching off electromagnetic interactions. No electrons are present in our system,
just the quark-gluon matter. In this system the
nonvanishing mass of the strange quark does not break the SU(2)$_{ CF}$
global symmetry.
 This is an important difference between our results concerning
 the  confined monopoles and those
reported in \cite{fthdqcd3}, see below. The degree of relevance of this
{\em gedanken} dense QCD, and consequences that ensue,
 to actual experiments on the quark-gluon plasma is to be investigated.

The split (\ref{vuvs}) breaks SU(3)$_{CF}$ and lifts
some of the orientational modes of the non-Abelian string.
This can be described by a small potential on the modular space of the string, i.e.
a potential term in the model (\ref{sac}). In supersymmetric theories
the scalar potentials in the
CP($N-1$) models on the string world sheet
induced by the squark mass differences were
calculated in \cite{SYmon,HT2}. At the same time, for
dense QCD this problem  was addressed in \cite{fthdqcd3}.

To calculate this potential to the leading order in
the small parameter $\epsilon$ in (\ref{potcorr}) we can still use the
solution (\ref{nastr}) where the profile functions are unchanged.
The only modification is that the common value of the scalar profile
functions at $r\to \infty$ in (\ref{bcinfty}) should be modified
as
\beq
v\to\tilde{v}, \qquad \tilde{v}^2=\frac{m_0^2-\frac23\epsilon}{8\lambda}\,,
\label{avv}
\eeq
which is the average value for the three VEVs in (\ref{qvevdef}).
The potential $(V+\delta V)$ gives the tension of the string
$2\pi\, \tilde{v}^2 \, \ln{(L m_0)}$ plus a finite  (nonlogarithmic) contribution
which depends on the moduli fields $n^{A}$. It is  given by
\beq
\frac{\epsilon}{3}\int d^4 x \,{\rm Tr}\left[
\Phi^{\dagger}\, {\rm diag}(1,1,-2)\, \Phi\right].
\eeq
Now, substituting here   Eq.~(\ref{nastr})  we get the potential in the
deformed CP(2) model on the string world sheet, to the leading order in $\epsilon/m_0^2$,
\beq
V_{ CP}= \omega \int d t\,dx_3\left(3n_3^2-1\right),
\label{CPpot}
\eeq
where
\beq
\omega= \frac{2\pi}{3}\,\epsilon\,\int_0^{\infty} rdr\left(\phi_2^2-\phi_1^2\right)\sim
\epsilon\,\frac{v^2}{m_0^2}\,,
\label{omega}
\eeq
Here we used the expansions (\ref{profileinfty}) and (\ref{profileinter}) to make the last estimate.

From (\ref{CPpot}) it is clearly seen  that with $m_s\neq 0$  the $(0,0,1)$ string has
a significantly larger tension than the $(1,0,0)$ and $(0,1,0)$ strings and is, in fact,
classically unstable. It is not even a local minimum of the potential (rather, it corresponds
to a maximum). Note that
the parameter $\omega$ is much larger than the quantum scale
$\Lambda_{CP}$ of the CP(2) model, a crucial circumstance.
Therefore, the classical splitting by far  dominates  over the quantum one in
(\ref{split}).

This instability  means, in particular,  that the monopole-antimonopole meson
formed through the insertion of a piece of the $(0,0,1)$ string in the
$(1,0,0)$ or  $(0,1,0)$ strings (see  Fig. \ref{kinkantikin}) is highly unstable and decays into
a perturbative state with the same global (singlet or adjoint) quantum numbers
with respect to the unbroken ${\rm SU}(2)_{CF}\times {\rm U}(1)$. This corresponds to the
process in which monopoles with the magnetic charges
given in (\ref{othermon}) are annihilated with their antimonopole
partners inside the monopole-antimonopole mesons. Correspondingly,
the monopoles with these magnetic charges disappear from the string.

The potential in (\ref{CPpot}) shows that the $n^3$ field  is heavy
and can be integrated out from the $CP(2)$ model under consideration. Then we are left
with the $CP(1)$ model (\ref{sac})
on the string world sheet,
which includes now only the fields $n^1$ and $n^2$, and no potential
on the target space. Its global group SU(2)$_{CF}$ remains unbroken.

In the quantum regime two non-Abelian strings whose low-energy dynamics is described
by this $CP(1)$  model (the $CP(1)$ vacua) are split, as was discussed in Sec. \ref{kkmesons},
see (\ref{split}). There are mesons on the strings with the lowest tension
which include pieces of the excited string. These are
formed by monopoles and
antimonopoles with magnetic charges classically given by
(\ref{12mon}). (Remember, in the quantum non-Abelian regime
the magnetic monopole charge  is averaged to zero.) Stable mesons are triplets
with respect to unbroken SU(2)$_{ CF}$.
Thus, our conclusion on the presence of the confined non-Abelian monopoles
attached to the non-Abelian strings in dense QCD stays {\em valid} even in a more realistic setting
of dense QCD with $N=3$ and
nonvanishing strange quark mass.

To conclude this section let us compare our results with those obtained in \cite{fthdqcd3}.
In \cite{fthdqcd3} a realistic
dense matter inside neutron stars was studied. In particular,
the electromagnetic interactions and the presence of electrons were
taken into account. As was mentioned above, this leads to
the complete breaking of the non-Abelian color-flavor symmetry
${\rm SU}(3)_{ CF}\to {\rm U}(1)^3$, All three strings are classically split by
the strange quark mass. Two excited
strings become classically unstable, and the monopoles effectively
disappear from the string. They are annihilated by the would-be antimonopoles.

In our paper we do not attempt to study realistic
  neutron stars. Instead, we focus just on the quark-gluon matter in   dense QCD,
   and demonstrate that in the CFL phase there are confined
   magnetic monopoles attached to the non-Abelian strings.
   Whether or not the quark-gluon plasma can exist in terrestrial
   experiments sufficiently long allowing for the
   formation of long non-Abelian strings is a separate issue
   left for further studies.

\subsection{Non-zero $u$ and $d$-quark masses}
\label{secmud}

Now let us  introduce nonvanishing $u$ and $d$ quark masses,
\beq
m_u=m_d\ll m_s.
\label{mud}
\eeq
We stress that we assume that  $u$ and $d$ quarks are  strictly
degenerate. This ensures that the color-flavor group
(\ref{newc+f}) remains unbroken.
The introduction of the common $u$ and $d$ quark mass just shifts
the parameter $m_0$ (and $\lambda$) in the GL model (\ref{glmodel})
 leaving  our results intact.

Another effect due to  $m_u=m_d\neq 0$  is that now
``pions'' become massive.
Their  masses were
estimated   (see e.g. the review  \cite{2}),
\beq
m_{\pi}\sim \sqrt{m_u m_s}\;\frac{T_c}{\mu}\,.
\label{mpi}
\eeq
For reasonable values of the quark masses  the ``pion'' masses are rather small,
$\sim m_{u,d} \,(6T_c/\mu )$. In particular, they are smaller
than the Higgsed gluon mass constrained by the week coupling condition (\ref{wccondition}).
Strictly speaking, this means that we cannot totally
ignore ``pions'' in our GL effective description (\ref{glmodel})
of  QCD in the CFL phase. They
 should be included in consideration of the low-energy theory.

The detailed study of the impact of these ``pions'' on
the non-Abelian strings is left for a future work.
Here we will make a few qualitative comments. First, they  are ``neutral"
with respect to the gauge fields and,
therefore, their presence in the bulk theory does not affect the classical solution (\ref{nastr}) for
the non-Abelian strings.
However, they will show up in loops producing, generally  speaking,  long-range tails
of string profile functions.
Effectively their presence forces the string to swell in the transverse dimension, acquiring the transverse size
of the  order of
$1/m_{\pi}$. We assume, however,  that
\beq
m_{\pi}\gg \Lambda_{CP}.
\label{mpiL}
\eeq
This constraint can be easily achieved since $\Lambda_{CP}$ is quite small, see (\ref{lambdaCP}).

The condition (\ref{mpiL}) ensures that the CP(1) model we arrived at
still can be used for the low-energy description of
dynamics of the
orienational zero modes on the non-Abelian string.
It means that the inverse transverse size of the string
(although small) is still much larger than typical
excitation energies on the world sheet, which are of
order of $\Lambda_{CP}$. Higher-derivative corrections to the CP(1) model
(see (\ref{sac}) with $N=2$)
run in powers of the ratio of the typical excitation energies over $m_\pi$
which can be considered  small due to the condition (\ref{mpiL}).
Of course, here we speak only about the rotational moduli fields
which are responsible for nearly degenerate strings and kinks/monopoles.

To conclude this section, we stress again that with   $m_u=m_d$ the color-flavor locked
SU(2)  stays unbroken. Therefore, two non-Abelian
strings described by CP(1) model are in highly  quantum regime which entails with necessity
 monopole-antimonopole pairs in the form of ``mesons"
attached to these strings.

\section{On the $\theta$ dependence}
\label{thetad}
\setcounter{equation}{0}

In this section we add the bulk $\theta$ term
and trace its impact on the
non-Abelian strings and monopoles of dense QCD.

\subsection{$\theta$ term on the world sheet}
\label{ttws}

In this section we will make a few comments concerning
possible effects due to the $\theta$ term and axions. In QCD all quarks
are   massive, hence,  the $\theta$ term effect
 cannot be eliminated by  chiral rotations. We are
interested in weather or not the $\theta$ term affects the world sheet theory
on the non-Abelian string in dense QCD.
The nonsupersymmetric  $CP(2)$ model allows one to introduce
a $\theta$ term which, as usual, is coupled to the topological
charge,
\beq
{\cal L}_\theta = \frac{\theta}{2\pi }\, \varepsilon_{\mu\nu} \partial^\mu A^\nu
=\frac{\theta}{2\pi }\, \varepsilon_{\mu\nu} \, \partial^\mu \left(
\bar{n}_i\partial^\nu n^i
\right)\,.
\label{two}
\eeq
Previously, in  the simplest model \cite{gsy}
we demonstrated
that the four-dimensional (bulk)
$\theta$ term penetrates
in  the two-dimensional sigma model,
\beq
\theta_{3+1}=\theta_{1+1}\,,
\label{theta}
\eeq
with self-evident notation.
 The above equality between the
 four- and two-dimensional $\theta$'s, however, is  {\em not}
a common property of all  non-Abelian strings, see  \cite{auzzi} for
a counterexample. To find out what happens in
 the  case at hand we can substitute the gauge field from
the string solution into the four-dimensional topological term
and integrate over the transverse directions.
As a result,  we get that the equality (\ref{theta}) is fulfilled.
There are nontrivial  phenomena in the
bulk theory  at $\theta=\pi$ due to vacuum double degeneracy.
On the
world sheet as well something remarkable occurs at $\theta =\pi$,
namely the {\em deconfinement} phenomenon. At this point the vacua   in the
and world sheet become  doubly degenerate too, and the single kink-monopole
state
becomes liberated and free to move along the string.

It is usually assumed that the $\theta$ dependence in
dense QCD is negligible  since the instanton-induced effects are
exponentially  suppressed due to   Higgsing and large values of the diquark condensates.
Our discussion
implies that the non-Abelian string provides a nontrivial environment
for the $\theta$ dependence to show up in full.

Care should be taken of the fermion modes
on the string. If there were  fermion zero modes
in the world sheet theory,  the   $\theta_{1+1}$
term could be eliminated from physics by   chiral rotations, as it happens in the supersymmetric version.
If the bulk quarks are strictly massless  the fermion
zero modes on the string do indeed exist \cite{nittafer}. In this limit
there is no $\theta$ dependence on the string.
However,  if non-zero  quark masses  are introduced,
the fermion zero modes on the string are lifted and the fermions become irrelevant.

Let us comment on the interaction of the non-Abelian
string in the CFL phase with the background fields.
The simplest example of the bulk field coupled to the string
is the axion. The
 string-axion interaction was considered
in \cite{axion} where it was argued that a kind of axion halo
emerges around the string, provided the string exists long enough
for the halo to form. For shorter time intervals the
string with kinks on it acts as an antenna for the axion emission.

Note also that due to the Witten effect monopoles acquire the electric
charge if the $\theta$ term is switched on. Hence the dyons emerge
and the monopole-antimonole pair gets substituted by the dyon-antidyon
state on the string worldsheet.

\subsection{A Holographic viewpoint}
\label{holovp}

Remembering that a holographic representation in the (nonsupersymmetric) problem at hand
can be exploited, if at all,
only for a general guidance,
let us have a closer look at  the derivation of the world-sheet
$\theta$ term via holography. Note  that the very
idea to look for such holographic realization of the
string and confined monopoles is that it could help
in interpretation of their hypothetical counterparts in different limits
of the full QCD phase diagram.

To this end,
we first comment on  the known dual representation of the non-Abelian strings.
In   supersymmetric QCD with the Fayet--Iliopoulos  term $\xi\neq 0$ the non-Abelian
string is represented by a D2 brane stretched between two
NS5 branes displaced by distance $\xi$ in some internal coordinate
\cite{HT1}. Evidently, the tension of the string
in the four-dimensional space-time is proportional to this distance. Similarly,
the monopoles confined on the string are represented
by the D2 branes with two internal world-volume coordinates \cite{HT1}.

Such non-Abelian strings as the D2 branes have a very clear-cut counterpart
in  holographic QCD with vanishing chemical potential  \cite{gv1,gv2}.
At zero temperature the dual geometry has the vanishing circle
D2 brane that can wrap around, which yields a very small tension
of the emerging magnetic string. Above the confinement-deconfinement phase transition,
which corresponds to the Hawking--Page transition on the dual side,
the  magnetic string acquires a nonvanishing tension,  since
the relevant vanishing circle disappears.

From the consideration above we have learned two
well-established facts concerning non-Abelian strings in dense QCD:
its tension is proportional to the diquark condensate squared
and it carries a nontrivial $\theta$ term on the world sheet.
These features have to be reproduced holographically.
Since we   deal  with conventional (nonsupersymmetric) QCD,  the best
we can do is to work with   the Sakai--Sugimoto model
\cite{Sakai:2004cn} which is a version of the black-hole
background \cite{wittentem}
and {\em qualitatively} seems to reproduce basic  QCD phenomena.
It involves in the thermal case two periodic
coordinates $x_5,t_E$, radial coordinate $U$ and the
internal $S^4$ manifold. Below the phase transition
the $(x_5,U)$ coordinates provide the cigar-like geometry,
while above $T_c$ it is the $(t_E,U)$ pair that yields this cigar-like geometry.

Let us try to argue that within the Sakai--Sugimoto
type models the best candidate for the non-Abelian string  is a
wrapped D6 brane in the geometry of the charged
black hole,  much in the same spirit as in  \cite{auzzi}.
The black hole charge corresponds to
density in the holographic picture. The D2 realization
cannot reproduce the correct string tension; hence, a
higher-dimensional D brane has to be involved. If there
are no $S^2$ cycles in the background there is no simple
possibility to get the correct tension from the D4 brane.
However, we cannot fully exclude the latter option
with a more contrived background.

Consider
the candidate D6 brane wrapped around $S^4$ and extended
along $x_5$. The Chern--Simons term on its world volume reads as
\beq
S_{CS}=\int d^7x \,C_0 \wedge F \wedge F \wedge F\,.
\eeq
and the integral over $S^4$ yields the factor $N$
amounting to
\beq
S_{CS}= N  \int dx_5 \, C_0 \int d^2 x \, F
\eeq
where $C_0\propto \theta_{3+1}$

If we assume that the integration runs over the whole $x_5$
circle we get a contradiction with the field-theoretical calculation
because of the $N$ factor.
To avoid this contradiction we assume that the integration runs
over the segment of the $x_5$ circle between two
flavor branes involved in the diquark condensate, which
yields an additional $\frac{1}{N_f}=\frac{1}{N }$ factor.
The tension of the string is given by
the area of the corresponding disc,  by the same token as in \cite{auzzi}.
Hence we can qualitatively reproduce the required features.
However, since there is no  clear-cut holographic
representation of the  CFL phase of dense QCD in the Sakai--Sugimoto model yet,
this D6 interpretation certainly calls for an additional  study.

\section{Conclusions}
\label{conccc}

What has been achieved in this work?
We started from the earlier observation of
non-Abelian strings   in the
color-flavor locked phase of dense QCD below $T_c$.
These strings develop orientational zero modes, which become
dynamical  fields of the $CP(2)$ model on the string world sheet.
The above model supports kinks (antikinks) which are confined to the string, and, moreover,
confined into kink-antikink bound states along the string,
albeit the kink constituents are still identifiable.

The most nontrivial part of our further argument is as follows.
We show that the kinks  appearing in the world-sheet theory on these strings,
in the form of the kink-antikink bound pairs, are, in fact,  magnetic monopoles,
as they manage to adapt and survive  in such a peculiar  form in dense QCD.  The kinks of $CP(2)$  are the descendants
of the 't Hooft--Polyakov monopoles --- the latter appear in the quasiclassical regime while the former
are the objects appearing in the highly quantum regime.
Our considerations are heavily based on analogies and inspiration we abstracted
from  certain supersymmetric non-Abelian theories.

This is the first ever analytic demonstration that the magnetic monopoles are   ``native"
to non-Abelian Yang--Mills theories such as QCD (albeit our analysis extends only to the phase of the monopole
confinement and has nothing to say about their condensation).
Abundant speculations can be presented here, but we will refrain from them at this stage
in the hope that a solid consideration allowing one to move
towards the monopole condensation
can be worked out later.

In conclusion,
let us comment on  possible signatures of the non-Abelian
strings in the neutron stars. The baryon chemical potential
in these stars depends on the location of the domain under consideration,
 and increases towards
the star center. In other words, the CFL phase can be realized
in a certain domain at a certain distance
 from the star center.
The non-Abelian strings can be created via rotations
of the neutron star \cite{Sedrakian:2008aya}.

The key question concerns   specific detectable signals from
the non-Abelian strings.
Being created by some mechanism, the non-Abelian string could emit
axions from   inside the star. Another
question which can be raised concerns the moving string.
Assume that the string  moves towards the boundary of the
star from the domain inside the star with the CFL phase. At a certain distance from the center
 the CFL phase becomes impossible, i.e.   the
non-Abelian string
solution is no longer supported. This means that at this point
the non-Abelian strings have to somehow join into the Abelian string excitations
existing in the subsequent 2SC phase.

Finally, we  would like to mention a very recent paper on non-Abelian strings in
dense QCD \cite{Nittaint} where interactions of string degrees of freedom with
light bulk fields are studied.

\section*{Acknowledgments}
This work was completed during an extended visit of MS to the MIT Center for Theoretical Physics.
MS would like to thank my colleagues for hospitality. The work of MS was supported in part by DOE
grant DE-FG02-94ER408. The work of AG is supported in part by the grants
RFBR-09-02-00308 and CRDF -  RUP2-2961-MO-09. AG thanks IPhT at Saclay
where a part of the work was carried out, for  hospitality and support.
The work of AY was  supported
by  FTPI, University of Minnesota,
by RFBR Grant No. 09-02-00457a
and by Russian State Grant for
Scientific Schools RSGSS-65751.2010.2.



\vspace{2.5cm}

\small
\begin{thebibliography}{99}
\itemsep -2pt

\bibitem{SW1}
N.~Seiberg and E.~Witten,
  Nucl.\ Phys.\  B {\bf 426}, 19 (1994)
  [Erratum-ibid.\  B {\bf 430}, 485 (1994)]
  [arXiv:hep-th/9407087];
  Nucl.\ Phys.\  B {\bf 431}, 484 (1994)
  [arXiv:hep-th/9408099].

\bibitem{HT1}
A.~Hanany and D.~Tong,
JHEP {\bf 0307}, 037 (2003)
[hep-th/0306150].

\bibitem{ABEKY}
R.~Auzzi, S.~Bolognesi, J.~Evslin, K.~Konishi and A.~Yung,
Nucl.\ Phys.\ B {\bf 673}, 187 (2003)
[hep-th/0307287].

\bibitem{SYmon}
M.~Shifman and A.~Yung,
Phys.\ Rev.\ D {\bf 70}, 045004 (2004)
[hep-th/0403149].

\bibitem{HT2}
A.~Hanany and D.~Tong,
JHEP {\bf 0404}, 066 (2004)
[hep-th/0403158].

\bibitem{fthdqcd2}
E.~Nakano, M.~Nitta and T.~Matsuura,
  Prog.\ Theor.\ Phys.\ Suppl.\  {\bf 174}, 254 (2008)
  [arXiv:0805.4539 [hep-ph]].

\bibitem{fthdqcd3}
M.~Eto, M.~Nitta and N.~Yamamoto,
  Phys.\ Rev.\ Lett.\  {\bf 104}, 161601 (2010)
  [arXiv:0912.1352 [hep-ph]].

\bibitem{gm}
  A.~Gorsky and V.~Mikhailov,
  Phys.\ Rev.\  D {\bf 76}, 105008 (2007)
  [arXiv:0707.2304 [hep-th]].

\bibitem{fthdqcd1}
A Yung, Unpublished, 2009.

\bibitem{Trev}
D.~Tong,
{\em TASI Lectures on Solitons,}
  arXiv:hep-th/0509216.

\bibitem{Jrev}
  M.~Eto, Y.~Isozumi, M.~Nitta, K.~Ohashi and N.~Sakai,
  J.\ Phys.\ A  {\bf 39}, R315 (2006)
  [arXiv:hep-th/0602170];
K.~Konishi,
  Lect.\ Notes Phys.\  {\bf 737}, 471 (2008)
  [arXiv:hep-th/0702102];

\bibitem{SYrev}
M.~Shifman and A.~Yung,
{\sl Supersymmetric Solitons,}
Rev.\ Mod.\ Phys. {\bf 79} 1139 (2007)
[arXiv:hep-th/0703267]; an expanded version in {\sl Supersymmetric Solitons,}
Cambridge University Press, Cambridge, 2009.

\bibitem{Trev2}
D.~Tong,
  Annals Phys.\  {\bf 324}, 30 (2009)
  [arXiv:0809.5060 [hep-th]].

\bibitem{1}
  K.~Rajagopal and F.~Wilczek,
{\em The condensed matter physics of QCD}, in {\sl At the Frontier of Particle Physics},
Ed. M. Shifman (World Scientific, 2001), Vol. 3, p. 2061
 [arXiv:hep-ph/0011333].

\bibitem{2}
M.~G.~Alford, A.~Schmitt, K.~Rajagopal and T.~Sch\"afer,
  Rev.\ Mod.\ Phys.\  {\bf 80}, 1455 (2008)
  [arXiv:0709.4635 [hep-ph]].

\bibitem{Shifman:2009mb}
  M.~Shifman and A.~Yung,
  Phys.\ Rev.\  D {\bf 79}, 125012 (2009)
  [arXiv:0904.1035 [hep-th]].

\bibitem{2SC}
  M.~G.~Alford, K.~Rajagopal and F.~Wilczek,
  Phys.\ Lett.\  B {\bf 422}, 247 (1998)
  [arXiv:hep-ph/9711395 ];
  R.~Rapp, T.~Sch\"afer, E.~V.~Shuryak and M.~Velkovsky,
  Phys.\ Rev.\ Lett.\  {\bf 81}, 53 (1998)
  [arXiv:hep-ph/9711396];
   B.~C.~Barrois,
{\em Nonperturbative Effects In Dense Quark Matter,}
PhD Thesis, 1979, CalTech UMI 79-04847;
 D.~Bailin and A.~Love,
  Phys.\ Rept.\  {\bf 107}, 325 (1984).

\bibitem{CFL}
  M.~G.~Alford, K.~Rajagopal and F.~Wilczek,
  Nucl.\ Phys.\ B {\bf 537} (1999) 443
  [arXiv:hep-ph/9804403].

\bibitem{ss}
  D.~T.~Son and M.~A.~Stephanov,
  Phys.\ Rev.\  D {\bf 61}, 074012 (2000)
  [arXiv:hep-ph/9910491];
  Erratum,
  Phys.\ Rev.\  D {\bf 62}, 059902 (2000)
  [arXiv:hep-ph/0004095].

\bibitem{GLdesc}
I.~Giannakis and H.~c.~Ren,
  Phys.\ Rev.\  D {\bf 65}, 054017 (2002)
  [arXiv:hep-ph/0108256];
K.~Iida and G.~Baym,
  Phys.\ Rev.\  D {\bf 63}, 074018 (2001)
  [Erratum-ibid.\  D {\bf 66}, 059903 (2002)]
  [arXiv:hep-ph/0011229].

\bibitem{BarH}
K.~Bardakci and M.~B.~Halpern,
Phys.\ Rev.\ D {\bf 6}, 696 (1972).

\bibitem{GiannRen}
I.~Giannakis, H.-c.~Ren,
Nucl. \ Phys. \  {\bf B669}, 462  (2003)
[hep-ph/0305235].
	
\bibitem{gsy}
  A.~Gorsky, M.~Shifman and A.~Yung,
  Phys.\ Rev.\  D {\bf 71}, 045010 (2005)
  [arXiv:hep-th/0412082].
	
\bibitem{EtoNitta}
M.~Eto and  M.~Nitta,
Phys. \ Rev. \ D{\bf 80} 125007 (2010)
[arXiv:0907.1278].

\bibitem{Y99}
A.~Yung,
 Nucl. \ Phys. \ B {\bf 562},   191  (1999)
[hep-th/9906243].

\bibitem{mat}
  A.~P.~Balachandran, S.~Digal and T.~Matsuura,
  Phys.\ Rev.\  D {\bf 73}, 074009 (2006)
  [arXiv:hep-ph/0509276].

\bibitem{iida}
   K.~Iida and G.~Baym,
  Phys.\ Rev.\  D {\bf 66}, 014015 (2002)
  [arXiv:hep-ph/0204124];
  K.~Iida,
  Phys.\ Rev.\  D {\bf 71}, 054011 (2005)
  [arXiv:hep-ph/0412426].

\bibitem{forbes}
  M.~M.~Forbes and A.~R.~Zhitnitsky,
  Phys.\ Rev.\  D {\bf 65}, 085009 (2002)
  [arXiv:hep-ph/0109173].

\bibitem{nittaws}
M.~Eto, E.~Nakano and  M.~Nitta,
	Phys. \ Rev. \ D {\bf 80} 125011 (2009)
	[arXiv:0908.4470].
	
\bibitem{AddaS}
  A.~D'Adda, M.~L\"uscher and P.~Di Vecchia,
  Nucl.\ Phys.\  B {\bf 146}, 63 (1978);
  A.~D'Adda, P.~Di Vecchia and M.~L\"uscher,
  Nucl.\ Phys.\  B {\bf 152}, 125 (1979).

\bibitem{WitS}
  E.~Witten,
  Nucl.\ Phys.\  B {\bf 149}, 285 (1979).

  \bibitem{wittenproved}
E.~Witten,
Phys.\ Rev.\ Lett.\  {\bf 81}, 2862 (1998)
[hep-th/9807109].

\bibitem{HoVa}
  K.~Hori and C.~Vafa,
{\em Mirror symmetry,}
  arXiv:hep-th/0002222.

  \bibitem{zamo}
   A.~B.~Zamolodchikov and A.~B.~Zamolodchikov,
  Annals Phys.\  {\bf 120}, 253 (1979).
  	
\bibitem{Tong}
D.~Tong,
Phys.\ Rev.\ D {\bf 69}, 065003 (2004)
[hep-th/0307302].

\bibitem{mmodel}
A.~Gorsky, M.~Shifman and A.~Yung,
Phys.\ Rev.\ D {\bf 75}, 065032 (2007)
[hep-th/0701040].

\bibitem{SYhetN}
  M.~Shifman and A.~Yung,
  Phys.\ Rev.\  D {\bf 77}, 125017 (2008)
  [arXiv:0803.0698 [hep-th]].

  \bibitem{Dorey}
N.~Dorey,
JHEP {\bf 9811}, 005 (1998) [hep-th/9806056].

\bibitem{HaHo}
A.~Hanany and K.~Hori,
  Nucl.\ Phys.\  B {\bf 513}, 119 (1998)
  [arXiv:hep-th/9707192].
[hep-th/9211097].

\bibitem{SYtorkink}
M.~Shifman and A.~Yung,
  Phys.\ Rev.\  D {\bf 81}, 085009 (2010)
  [arXiv:1002.0322 [hep-th]].

\bibitem{Iidams}
K.~Iida, T.~Matsuura, M.~Tachibana and  T.~Hatsuda,
Phys. \ Rev. \ Lett. {\bf 93} (2004) 132001
[arXiv:hep-ph/0312363].

\bibitem{auzzi}
  R.~Auzzi and S.~Prem Kumar,
  JHEP {\bf 0812}, 077 (2008)
  [arXiv:0810.3201 [hep-th]];
  Phys.\ Rev.\ Lett.\  {\bf 103}, 231601 (2009)
  [arXiv:0908.4278 [hep-th]].

\bibitem{nittafer}
  S.~Yasui, K.~Itakura and M.~Nitta,
  Phys.\ Rev.\  D {\bf 81}, 105003 (2010)
  [arXiv:1001.3730 [math-ph]].

\bibitem{axion}
  A.~Gorsky, M.~Shifman and A.~Yung,
  Phys.\ Rev.\  D {\bf 73}, 125011 (2006)
  [arXiv:hep-th/0601131].

\bibitem{gv1}
  A.~Gorsky and V.~Zakharov,
  Phys.\ Rev.\  D {\bf 77}, 045017 (2008)
  [arXiv:0707.1284 [hep-th]].

 \bibitem{gv2}
 A.~S.~Gorsky, V.~I.~Zakharov and A.~R.~Zhitnitsky,
  Phys.\ Rev.\  D {\bf 79}, 106003 (2009)
  [arXiv:0902.1842 [hep-ph]].

  \bibitem{Sakai:2004cn}
  T.~Sakai and S.~Sugimoto,
  Prog.\ Theor.\ Phys.\  {\bf 113}, 843 (2005)
  [arXiv:hep-th/0412141].
  
\bibitem{wittentem}
  E.~Witten,
  Adv.\ Theor.\ Math.\ Phys.\  {\bf 2}, 505 (1998)
  [arXiv:hep-th/9803131].

  \bibitem{Sedrakian:2008aya}
  D.~M.~Sedrakian, D.~Blaschke, K.~M.~Shahabasyan and M.~K.~Shahabasyan,
{\em Vortex structure of a neutron star with CFL quark core,}
  arXiv:0810.3003 [hep-ph].

\bibitem{Nittaint}
Y.~Hirono, T.~Kanazawa and  M.~Nitta,
{\em Topological Interactions of Non-Abelian Vortices with Quasi-Particles in High Density QCD,}
  arXiv:1012.6042 [hep-ph].




\end{thebibliography}
\end{document}